\documentclass[pra,twocolumn,showpacs,preprintnumbers,amsmath,amssymb]{revtex4}
\usepackage{graphicx}
\usepackage{dcolumn}
\usepackage{bm}
\usepackage{amsthm}
\usepackage{color}
\input xy
\xyoption{all}

\newcommand{\ket}[1]{\left| #1 \right\rangle}

\newcommand{\g}{\mathfrak{g}}

\newcommand{\C}{\mathbb{C}}
\newtheorem{theorem}{Theorem}
\newtheorem{lemma}{Lemma}
\newtheorem{proposition}{Proposition}
\newtheorem{definition}{Definition}
\newtheorem{conjecture}{Conjecture}
\newtheorem{corollary}{Corollary}

\DeclareMathOperator{\Lie}{Lie}
\DeclareMathOperator{\Stab}{Stab}

\begin{document}

\title{Maximum stabilizer dimension for nonproduct states}

\author{Scott N. Walck}
  \email{walck@lvc.edu}
\author{David W. Lyons}
  \email{lyons@lvc.edu}
\affiliation{Lebanon Valley College, Annville, PA 17003}

\date{2 August 2007}

\begin{abstract}
Composite quantum states can be classified by how they behave under
local unitary transformations.  Each quantum state has a stabilizer
subgroup and a corresponding Lie algebra,
the structure of which is a local unitary invariant.
In this paper, we study the structure of the stabilizer subalgebra
for $n$-qubit pure states,
and find its maximum dimension to be $n-1$
for nonproduct states of three qubits and higher.
The $n$-qubit Greenberger-Horne-Zeilinger
state has a stabilizer subalgebra that achieves the
maximum possible dimension for pure nonproduct states.
The converse, however, is not true:  we show examples of
pure 4-qubit states that achieve the maximum nonproduct
stabilizer dimension,
but have stabilizer subalgebra structures different from that
of the $n$-qubit GHZ state.
\end{abstract}


\maketitle

\section{Introduction}

Understanding multipartite entanglement is an important problem
in quantum information science.
One of the first and most natural proposals for classifying
entanglement was to regard two quantum states $\ket{\psi}$
and $\ket{\psi'}$
as having the same
type of entanglement if they are related by a local unitary
transformation,
\[
\ket{\psi'} = U_1 \otimes U_2 \otimes \cdots \otimes U_n \ket{\psi}
\]
where unitary operator $U_i$ acts on subsystem $i$
\cite{linden98,linden99}.
The local unitary group partitions the space of quantum states
into orbits, with each orbit representing a type of entanglement.
There is an intimate relationship between the orbit of a state
$\ket{\psi}$ and the stabilizer subgroup of $\ket{\psi}$, that is, the
subgroup of the local unitary group that leaves $\ket{\psi}$ invariant.
The orbit $\mathcal{O}_\psi$ of state $\ket{\psi}$
is the quotient space
of the local unitary group $G$ by the stabilizer subgroup
$\Stab{\psi}$.
\[
\mathcal{O}_\psi = G/\Stab(\psi)
\]
Studying stabilizer subgroups is therefore equivalent in some sense to
studying local unitary orbits.

The full classification problem for multipartite entanglement
has been a very difficult problem,
and complete results exist only for 2- and 3-qubit pure states
\cite{carteret00a,sudbery01,acin00,acin01,walck3qubit}.
Progress has been made, however, in understanding aspects
of the local unitary orbits and their dimensions
\cite{carteret00,carteret00a,kus01,sinolecka02,lyonswalck1,lyonswalck2}.

In two previous papers \cite{lyonswalck1,lyonswalck2},
we showed that that maximum stabilizer dimension for $n$-qubit
pure states is $3n/2$ (for $n$ even) and $(3n-1)/2$ (for $n$ odd), that
products of singlet pairs,
\[
\ket{\psi}
 = (\ket{01} - \ket{10}) \otimes \cdots \otimes (\ket{01} - \ket{10}) ,
\]
achieve
this maximum stabilizer dimension, and that such products of
singlets (and their LU equivalents)
are the \emph{only} pure $n$-qubit states to achieve
the maximum stabilizer dimension.

From the perspective of stabilizer structure, nonproduct states
are the basic components from which all pure states are built.
Product states $\ket{\psi} = \ket{\phi} \otimes \ket{\chi}$
simply inherit the structure of their factor states.
For this reason, it is appropriate to ask the fundamental questions
about stabilizer dimensions for nonproduct states.

In this paper, we consider pure $n$-qubit quantum states.
We study the structure of their
local unitary stabilizer subalgebras, and apply the structure
results to prove that the maximum stabilizer dimension for
nonproduct states is $n-1$ (for $n \geq 3$).
We also achieve some previously reported results in a simpler
fashion.

Section \ref{notation} fixes the notation that we will use.
Section \ref{structure} gives our results concerning the
structure of $n$-qubit local unitary stabilizer subalgebras.
Section \ref{maxdim} applies the results of section \ref{structure}
to achieve the maximum stabilizer dimension for nonproduct states,
as well as some other maximum dimension results.
Section \ref{ncat} shows that the $n$-qubit GHZ state achieves the
maximum nonproduct stabilizer dimension, but that, at least in
the case of 4-qubit states, it is not the only such state.

\section{Notation}
\label{notation}

Let $H = (\C^2)^{\otimes n}$ be the $n$-qubit Hilbert space.
The local unitary group $G = U(1) \times SU(2)^n$ and its
Lie algebra $\g = \Lie(G) = LG$ have dimension $3n+1$.
We have
\[
\g = \g_0 \oplus \g_1 \oplus \cdots \oplus \g_n
\]
where $\g_0$ is one-dimensional and $\g_j = su(2)$
(the set of $2 \times 2$ skew-Hermitian matrices)
is three-dimensional for $j=1,\ldots,n$.
For a set of qubits $\mathcal{S} = \{j_1,\ldots,j_m\}$,
we write $\g_\mathcal{S} = \g_{j_1} \oplus \cdots \oplus \g_{j_m}$,
and
$\overline{\g_\mathcal{S}}
 = \g_0 \oplus \g_{\mathcal{N} \setminus \mathcal{S}}$,
where $\mathcal{N} = \{1,\ldots,n\}$.
When we need a basis for $su(2)$, we will use
\begin{align*}
A &= \left[ \begin{array}{cc} i & 0 \\ 0 & -i \end{array} \right] , &
B &= \left[ \begin{array}{cc} 0 & 1 \\ -1 & 0 \end{array} \right] , &
C &= \left[ \begin{array}{cc} 0 & i \\  i & 0 \end{array} \right] .
\end{align*}
The element $A_j \in \g$ for $1 \leq j \leq n$ has $A$
in the $j$th qubit slot and zeros elsewhere, and analogous notation
applies for $B_j$ and $C_j$.

The local unitary group action is a map
$\Phi: G \times H \to H$
which associates to each element $g$ of the local unitary group $G$
and each vector $\psi$ in the Hilbert space $H$ the vector
$\Phi(g,h) \in H$ obtained by acting on $\psi$ with the local
unitary transformation $g$.
If we pick a point (state vector) $\psi \in H$, we obtain a map
\[
\Phi_\psi: G \to H
\]
in which $\Phi_\psi(g) = \Phi(g,\psi)$ for $g \in G$.
The derivative of $\Phi_\psi$ at the identity is the map
\[
d\Phi_\psi: \g \to T_\psi H ,
\]
where $T_\psi H$ denotes the tangent space to $H$ at $\psi$,
and is naturally isomorphic with $H$.
Thus, we have a map
\[
d\Phi_\psi: \g \to H .
\]
We are particularly interested in the kernel of $d\Phi_\psi$,
which represents those elements of the local unitary Lie algebra
that send $\psi$ to zero.
We define $K_\psi = \ker d\Phi_\psi$ to be the stabilizer subalgebra
of the local unitary Lie algebra $\g$.
This $K_\psi$ is the Lie algebra of the stabilizer subgroup
$\Stab(\psi) \subset G$, the subgroup of local unitary
transformations that leave $\psi$ fixed.

We also define projection operators
\[
P_j: \g \to \g_j
\]
for each qubit $j$.

A multi-index $I = (i_1 \ldots i_n)$
is a sequence of $n$ binary digits
used to label one of the $2^n$
basis states of an $n$-qubit pure state vector.

\section{Structure of $K_\psi$}
\label{structure}

With the following definition and lemma we establish
that groups of qubits
can combine to form $su(2)$ summands in the stabilizer subalgebra.
We show that once they do so, those qubits are prohibited
from participating in the stabilizer subalgebra in any other way,
for example by contributing part of a one-dimensional summand
in $K_\psi$.

\begin{definition}
An $su(2)$ block for $\psi$ is a set
$\mathcal{S} \subset \{1,\ldots,n\}$ of qubits for which
\begin{enumerate}
\item[(a)] $\dim K_\psi \cap \g_\mathcal{S} > 1$, and
\item[(b)] every proper subset $\mathcal{S'} \subsetneq \mathcal{S}$
of qubits has
\[
\dim K_\psi \cap \g_\mathcal{S'} = 0 .
\]
\end{enumerate}
\end{definition}

\begin{lemma}
\label{kernsplit}
If $\mathcal{S}$ is an $su(2)$ block of $m$ qubits for $\psi$, then
\begin{enumerate}
\item[(a)] For each $l \in \{1,\ldots,m\}$, there are
$U_{j_l},V_{j_l},W_{j_l} \in \g_{j_l}$ with
\begin{align*}
[U_{j_l},V_{j_l}] &= W_{j_l}, & [V_{j_l},W_{j_l}] &= U_{j_l}, & [W_{j_l},U_{j_l}] &= V_{j_l}
\end{align*}
such that
\begin{align*}
U &:= \sum_{l=1}^m U_{j_l} \in K_\psi \\
V &:= \sum_{l=1}^m V_{j_l} \in K_\psi \\
W &:= \sum_{l=1}^m W_{j_l} \in K_\psi . \\
\end{align*}
\item[(b)] $K_\psi \cap \g_\mathcal{S} \cong su(2)$,
\item[(c)] $K_\psi = (K_\psi \cap \g_\mathcal{S}) \oplus (K_\psi \cap \overline{\g_\mathcal{S}})$.
\end{enumerate}
\end{lemma}
\begin{proof}
(a)
Suppose $\dim K_\psi \cap \g_\mathcal{S} > 1$.
Then $\g_\mathcal{S}$ contains at least two linearly independent elements
of $K_\psi$.
Let us call them
\[
X = \sum_{l=1}^m X_{j_l} \in K_\psi
\]
and
\[
Y = \sum_{l=1}^m Y_{j_l} \in K_\psi ,
\]
where $X_{j_l},Y_{j_l} \in \g_{j_l}$.
Define
\[
Z := [X,Y] = \left[ \sum_{l=1}^m X_{j_l} , \sum_{l=1}^m Y_{j_l} \right] = \sum_{l=1}^m [X_{j_l},Y_{j_l}]
           = \sum_{l=1}^m Z_{j_l} ,
\]
with $Z_{j_l} = [X_{j_l},Y_{j_l}]$.
This $Z \in K_\psi$, since $K_\psi$ is a Lie subalgebra of $\g$.
We know that $Z \neq 0$, since $X$ and $Y$ are linearly independent.
(This follows from the fact that two vectors in $su(2)$ have a trivial
bracket if and only if they are linearly dependent.)
Furthermore, we have $Z_{j_l} \neq 0$ for each $l$,
since otherwise $Z \in K_\psi \cap \g_\mathcal{S'}$
for some proper subset $\mathcal{S'} \subsetneq \mathcal{S}$ of qubits,
violating the definition of $su(2)$ block.
Since $Z_{j_l} \neq 0$ for each $l$,
we also know that
$X_{j_l} \neq 0$ and $Y_{j_l} \neq 0$ for each $l$,
and that
$X_{j_l}$ and $Y_{j_l}$ are linearly independent for each $l$.
We see that $X_{j_l}$, $Y_{j_l}$, and $Z_{j_l}$
span the 3 dimensions of $\g_{j_l}$ for each $l$.

Next define
\[
X' := [Y,Z] = \sum_{l=1}^m X'_{j_l} ,
\]
with $X'_{j_l} = [Y_{j_l},Z_{j_l}]$.
We have
\[
[Z,X'] = \sum_{l=1}^m [Z_{j_l},X'_{j_l}] = \sum_{l=1}^m b_l Y_{j_l} ,
\]
with $b_l > 0$ for each $l$,
since $\g_{j_l} = su(2)$ is isomorphic to the Lie algebra of three-dimensional real vectors
with cross-product as Lie bracket.
But if the $b_l$ are not the same for all $l$, then $[Z,X']$
is linearly independent of $Y$,
and a linear combination of $Y$ and $[Z,X']$
would produce a stabilizer subalgebra element in
$\g_\mathcal{S'}$ for some proper subset
$\mathcal{S'} \subsetneq \mathcal{S}$ of qubits.
We conclude that $[Z,X'] = b Y$,
where $b$ is the common value of the $b_l$ for $1 \leq l \leq n$.
With the same argument, we conclude that $[X',Y] = c Z$
for some $c > 0$.
Now let
\begin{align*}
U_{j_l} &= \frac{1}{\sqrt{b c}} X'_{j_l} &
V_{j_l} &= \frac{1}{\sqrt{c}} Y_{j_l} &
W_{j_l} &= \frac{1}{\sqrt{b}} Z_{j_l} ,
\end{align*}
and part (a) is proved.

(b) In part (a), we constructed 3 linearly independent vectors
in $K_\psi \cap \g_\mathcal{S}$, so
$\dim K_\psi \cap \g_\mathcal{S} \ge 3$.
But if $\dim K_\psi \cap \g_\mathcal{S} > 3$,
then there is a fourth element 
$R := \sum_{l=1}^m R_{j_l} \in K_\psi$.
But $R_{j_1}$ must be a linear combination of
$U_{j_1}$, $V_{j_1}$, and $W_{j_1}$, so there
is a linear combination of $U$, $V$, $W$, and $R$
making a stabilizer subalgebra element in
$\g_\mathcal{S'}$ for some proper subset
$\mathcal{S'} \subsetneq \mathcal{S}$ of qubits.

We conclude that
$K_\psi \cap \g_\mathcal{S}$ is a 3-dimensional Lie algebra
with basis $U$, $V$, $W$.  The bracket relations among the
three basis vectors establish that
$K_\psi \cap \g_\mathcal{S} \cong su(2)$.

(c)
Let $Q+Q' \in K_\psi$, where $Q \in \g_\mathcal{S}$ and $Q' \in \overline{\g_\mathcal{S}}$.
Then
\begin{align*}
[Q + Q',U] &= [Q,U] \in K_\psi, \\
[Q + Q',V] &= [Q,V] \in K_\psi, \\
[Q + Q',W] &= [Q,W] \in K_\psi .
\end{align*}
Let
\[
Q = \sum_{l=1}^m \alpha_l U_{j_l} + \sum_{l=1}^m \beta_l V_{j_l} + \sum_{l=1}^m \gamma_l W_{j_l} .
\]
Then
\[
[Q,U] = -\sum_{l=1}^m \beta_l W_{j_l} + \sum_{l=1}^m \gamma_l V_{j_l} .
\]
But $[Q,U] \in K_\psi \cap \g_\mathcal{S}$, so $[Q,U]$ is a linear combination of $U,V,W$.
It follows that $\beta_l$ must be the same for all $l$, and similarly for $\gamma_l$.
Consideration of $[Q,V]$ shows that $\alpha_l$ must be the same for all $l$.
It follows that $Q$ is a linear combination of $U,V,W$, hence that $Q \in K_\psi$,
and that $Q' \in K_\psi$.
\end{proof}

The next proposition provides a convenient way to identify
whether or not a qubit participates in an $su(2)$ block.
Following that, we note that a qubit may participate in zero
or one $su(2)$ block, but no more.

\begin{proposition}
\label{kerproj}
Let $\psi \in H$.
Qubit $j$ belongs to an $su(2)$ block for $\psi$
if and only if
$\dim P_j K_\psi > 1$.
\end{proposition}
\begin{proof}
If qubit $j$ belongs to an $su(2)$ block for $\psi$,
then $\dim P_j K_\psi = 3$ by part (a) of
Lemma \ref{kernsplit}.

Suppose a qubit $j$ with $\dim P_j K_\psi > 1$.
Let $X_j,Y_j \in P_j K_\psi$ be linearly independent.
Notice that $\dim P_j K_\psi = 3$, since
$P_j [X,Y]$ is linearly independent of $P_j X$ and $P_j Y$
when $X,Y \in K_\psi$ are chosen so that $P_j X = X_j$ and $P_j Y = Y_j$.
Let us take $X_j,Y_j,Z_j \in P_j K_\psi = \g_j$ with
\begin{align*}
[X_j,Y_j] &= Z_j & [Y_j,Z_j] &= X_j & [Z_j,X_j] &= Y_j .
\end{align*}

Let $X \in K_\psi$ be an element that projects onto a minimal number of qubits
consistent with the constraint that $P_j X = X_j$.
Let $\mathcal{S}$ be the set of qubits onto which $X$ has nontrivial projection.
Let $Y \in K_\psi$ be an element that projects onto a minimal number of qubits
consistent with the constraint that $P_j Y = Y_j$.
Let $\mathcal{T}$ be the set of qubits onto which $Y$ has nontrivial projection.
Now,
$Z = [X,Y]$ (which is not zero, since $[X_j,Y_j] \neq 0$) projects
onto the intersection $\mathcal{S} \cap \mathcal{T}$.
Then $[Y,Z]$ projects onto $\mathcal{S} \cap \mathcal{T}$,
contradicting the minimality of $\mathcal{S}$ (note $P_j [Y,Z] = X_j$)
unless $\mathcal{S} \subset \mathcal{T}$.
Finally, $[Z,X]$ projects onto $\mathcal{S} \cap \mathcal{T}$,
contradicting the minimality of $\mathcal{T}$ (note $P_j [Z,X] = Y_j$)
unless
$\mathcal{T} \subset \mathcal{S}$.
We conclude $\mathcal{S} = \mathcal{T}$.

Since $X,Y \in K_\psi \cap \g_\mathcal{S}$ are linearly independent,
and since $\mathcal{S}$
is minimal, $\mathcal{S}$ fulfills the requirements of an $su(2)$ block.
\end{proof}

\begin{proposition}
Let $\psi \in H$.  A qubit $j$ belongs to at most one $su(2)$ block
for $\psi$.
\end{proposition}
\begin{proof}
If $\mathcal{S}$ is an $su(2)$ block containing $j$ and
$\mathcal{T}$ is an $su(2)$ block containing $j$, then
$\mathcal{S} \cap \mathcal{T}$ is an $su(2)$ block containing
$j$, which violates the definition of $su(2)$ block unless
$\mathcal{S} = \mathcal{T}$.
\end{proof}

We have at this point established much of the essential structure
of $K_\psi$ that we need to place limits on its size.
The following dimension formula summarizes our findings that
some number of qubits will participate in $su(2)$ blocks, and that
each $su(2)$ block contributes three dimensions to the stabilizer
subalgebra.

\begin{theorem}[Dimension formula]
Let $\psi \in H$ be an $n$-qubit state vector,
let $p$ be the number of $su(2)$ blocks for $\psi$,
and let $\mathcal{B}$ be the set of qubits occurring in $su(2)$ blocks.
Then
\[
\dim K_\psi = 3p + \dim \left( K_\psi \cap \overline{\g_\mathcal{B}} \right) .
\]
\end{theorem}
\begin{proof}
This follows from part (c) of Lemma \ref{kernsplit}.
\end{proof}

The following lemma and proposition
show that each $su(2)$ block must be composed of
an even number of qubits.

\begin{lemma}
Let $\psi \in H$.  If
\[
\sum_{l=1}^{m} A_{j_l} \in K_\psi ,
\]
then $m$ is even.
\end{lemma}
\begin{proof}
We have, from \cite{lyonswalck1} equation (10),
\[
A_k \ket{\psi} = \sum_I i (-1)^{i_k} c_I \ket{I} .
\]
Since $\sum_{l=1}^{m} A_{j_l} \in K_\psi$,
we have
\[
\sum_{l=1}^{m} A_{j_l} \ket{\psi}
 = \sum_{l=1}^{m} \sum_I i (-1)^{i_{j_l}} c_I \ket{I} = 0 .
\]
Choose a multi-index $I$ with $c_I \neq 0$.
Then we must have
\[
\sum_{l=1}^{m} (-1)^{i_{j_l}} = 0 .
\]
The sum of a number of positive and negative ones can only be zero
if the number is even.
\end{proof}

\begin{proposition}
An $su(2)$ block must contain an even number of qubits.
\end{proposition}
\begin{proof}
Let $\psi \in H$ be an $n$-qubit state vector.
Let $\mathcal{S} = \{j_1,\ldots,j_m\}$ be an $su(2)$ block
for $\psi$.  Part (a) of Lemma \ref{kernsplit} guarantees
a
\[
U = \sum_{l=1}^m U_{j_l} \in K_\psi .
\]
There is a $\psi' \in H$, LU-equivalent to $\psi$,
for which
\[
\sum_{l=1}^m A_{j_l} \in K_{\psi'} .
\]
By the previous Lemma, $m$ must be even.
\end{proof}

The next lemma and its corollary concern
the contribution
to $K_\psi$ from qubits that are \emph{not} contained in $su(2)$ blocks.

\begin{lemma}
\label{nonblock}
Let $\psi \in H$ be an $n$-qubit state vector, and let
$\mathcal{B} = \{j_1,\ldots,j_b\}$ be
the set of qubits occurring in $su(2)$ blocks.
Then
\[
\dim \left( K_\psi \cap \overline{\g_\mathcal{B}} \right) \leq n-b .
\]
\end{lemma}
\begin{proof}
Since $\dim P_j K_\psi \leq 1$ for each qubit $j$ not contained in an $su(2)$ block
(Proposition \ref{kerproj}),
the projection of $K_\psi$ onto $\overline{\g_\mathcal{B}}$
(which is equal to $K_\psi \cap \overline{\g_\mathcal{B}}$)
can be at most $n-b+1$ dimensional.
But if $\dim (K_\psi \cap \overline{\g_\mathcal{B}}) = n-b+1$,
then elements of $\g_0$
are in $K_\psi$, and this is impossible.
Therefore, the maximum dimension of
$K_\psi \cap \overline{\g_\mathcal{B}}$ is $n-b$.
\end{proof}

\begin{corollary}
Let $\psi \in H$ be an $n$-qubit state vector with $b$ qubits
belonging to $p$ $su(2)$ blocks.
Then
\[
\dim K_\psi \leq 3p + n - b .
\]
\end{corollary}

Until this point, our results about stabilizer subalgebra structure
have applied to general pure states (either product or nonproduct
states).
The last two results of the section lay the groundwork for
consideration of nonproduct states.

First, we record a result obtained previously that two-qubit
$su(2)$ blocks occur precisely when a state is a product
of a singlet (or LU-equivalent) and another state \cite{lyonswalck2}.

\begin{proposition}
Let $\psi \in H$ be an $n$-qubit state vector.
There is an $su(2)$ block for $\psi$
containing exactly 2 qubits
if and only if $\psi$ is the product of a
2-qubit singlet (or LU equivalent) and an $(n-2)$-qubit state vector.
\end{proposition}
\begin{proof}
This is implied by Proposition 3.7 of \cite{lyonswalck2}.
\end{proof}

Finally, we have a strengthened version of Lemma \ref{nonblock}
for a state without a single qubit factor.

\begin{lemma}
\label{nonblocknosingle}
Let $\psi \in H$ be an $n$-qubit state vector, and let
$\mathcal{B} = \{j_1,\ldots,j_b\}$ be
the set of qubits occurring in $su(2)$ blocks.
If $\psi$ does not contain a single qubit factor
and there are a positive number of qubits not contained in
$su(2)$ blocks ($b < n$),
then
\[
\dim \left( K_\psi \cap \overline{\g_\mathcal{B}} \right) \leq n-b-1 .
\]
\end{lemma}
\begin{proof}
If $b < n$ and
\[
\dim \left( K_\psi \cap \overline{\g_\mathcal{B}} \right) = n-b
\]
then $\g_0 \oplus \g_j$ contains an element
of $K_\psi$ for every qubit $j$ not in an $su(2)$ block.
Lemma 3.8 of \cite{lyonswalck2} then
implies that $\psi$ contains a single qubit factor
(in fact, $n-b$ of them).
We conclude that if $\psi$ contains no single qubit factor,
then $\dim \left( K_\psi \cap \overline{\g_\mathcal{B}} \right) < n-b$,
and the result follows.
\end{proof}

\section{Maximum Dimension Theorems}
\label{maxdim}

In this section, we apply the results from the previous section
about the structure of the stabilizer subalgebra to find limits
on its dimension.
The first theorem applies in a general setting, without yet
restricting our attention to nonproduct states.  This result
was originally reported in \cite{lyonswalck1}.

\begin{theorem}[Maximum stabilizer dimension]
Let $\psi \in H$ be an $n$-qubit state vector.
Then
\[
\dim K_\psi \leq \frac{3n}{2} .
\]
\end{theorem}
\begin{proof}
Let $b$ be the number of qubits belonging to $su(2)$ blocks, and
$p$ the number of $su(2)$ blocks.
We must have $b \leq n$.
At least 2 qubits must belong to each block, so $b \geq 2p$.
This gives
\[
\dim K_\psi \leq 3p + n - b \leq \frac{b}{2} + n \leq \frac{3n}{2} .
\]
\end{proof}

At this point we begin to confine our attention to nonproduct states.
Since products of singlets give the largest stabilizer dimension,
a first step along this path is to consider the maximum stabilizer
dimension for any state that does not have a singlet factor.
By ``singlet'' we mean singlet or LU-equivalent.

\begin{theorem}[Non-singlet maximum stabilizer dimension]
\label{thmnonsing}
Let $\psi \in H$ be an $n$-qubit state vector
that does not contain a singlet factor.
Then $\dim K_\psi \leq n$.
\end{theorem}
\begin{proof}
Let $b$ be the number of qubits belonging to $su(2)$ blocks, and
$p$ the number of $su(2)$ blocks.
We must have $b \leq n$.
Since there are no singlet factors, we must have
at least 4 qubits belonging to each block, so $b \geq 4p$.
This gives
\[
\dim K_\psi \leq 3p + n - b \leq n - p .
\]
\end{proof}

We can squeeze a bit more out of the previous result if, in addition
to disallowing a singlet factor, we also disallow a single qubit
factor.

\begin{theorem}
\label{maxdimnosingletnosingle}
Let $\psi \in H$ be an $n$-qubit state vector
that does not contain a singlet factor,
and that does not contain a single qubit factor.
Then $\dim K_\psi \leq n-1$.
\end{theorem}
\begin{proof}
If all qubits participate in $su(2)$ blocks, then Theorem \ref{thmnonsing}
applies, and we can conclude $\dim K_\psi \leq n-1$.
If at least one qubit does not participate in an $su(2)$ block,
then Lemma \ref{nonblocknosingle} applies, and we have
\begin{align*}
\dim K_\psi
 &= 3p + \dim \left( K_\psi \cap \overline{\g_\mathcal{B}} \right) \\
 &\leq 3p + n - b - 1 \\
 &\leq n - p - 1 ,
\end{align*}
from which the result follows.
\end{proof}

Eliminating the possibility of one-qubit factors and two-qubit
singlet factors is already enough to conclude the main result.

\begin{corollary}[Nonproduct maximum stabilizer dimension]
\label{npmaxstabdim}
Let $\psi \in H$ be an $n$-qubit state vector
that does not factor into a product of an $m$-qubit state
and a $(n-m)$-qubit state for any choice of $m < n$ qubits.
Then
\[
\dim K_\psi \leq \left\{ \begin{array}{lll}
1 & , & n = 1 \\
3 & , & n = 2 \\
n-1 & , & n \geq 3
\end{array} \right. .
\]
\end{corollary}

\section{Generalized $n$-qubit GHZ states}
\label{ncat}

The generalized $n$-qubit GHZ state ($\alpha \neq 0$, $\beta \neq 0$)
\[
\ket{\psi} = \alpha \ket{00\dots 0} + \beta \ket{11\dots 1}
\]
has (for $n \geq 3$) stabilizer subalgebra
\[
K_\psi = \left\{ \sum_{j=1}^n t_j A_j \left| \sum_{j=1}^n t_j = 0
          \right. \right\} .
\]
It is easy to see that every element of the right hand side kills $\psi$
(since $A_j \ket{\psi} = A_k \ket{\psi}$ for every $j,k$),
and that it has dimension $n-1$.
By Corollary \ref{npmaxstabdim}, it must constitute the entire stabilizer
subalgebra.
We record this as a proposition.

\begin{proposition}
For $n \geq 3$, the generalized $n$-qubit GHZ state
has $\dim K_\psi = n - 1$.
\end{proposition}

The table below gives stabilizer subalgebra dimensions for a number of
pure $n$-qubit states, along with maximum stabilizer dimensions
in various circumstances.

\begin{center}
\begin{tabular}{lc}
      & Stabilizer \\
State & dimension \\ \hline
Generic $n$-qubit state, $n \geq 3$ & 0 \\
All 1-qubit states & 1 \\
Generic 2-qubit states & 1 \\
2-qubit product states & 2 \\
2-qubit singlet state  & 3 \\
Unentangled states     & $n$ \\
Product of singlets, $n$ even & $3n/2$ \\
Product of singlets, $n$ odd & $(3n-1)/2$ \\
$n$-qubit GHZ, $n \geq 3$   & $n-1$ \\ \hline
Maximum stabilizer dimension, $n$ even & $3n/2$ \\
Maximum stabilizer dimension, $n$ odd & $(3n-1)/2$ \\
Max. nonproduct stab. dimension, $n \geq 3$ & $n-1$
\end{tabular}
\end{center}

It is natural to ask at this point whether the results on stabilizer
dimensions for nonproduct states parallel those for general pure states.
In the general case, products of singlets and their LU-equivalents
achieve the maximum stabilizer dimension, and they are the \emph{only}
pure states to achieve that dimension.  We have seen that generalized
$n$-qubit GHZ states achieve the maximum nonproduct stabilizer dimension,
and it is natural to ask whether they are the only nonproduct states
to achieve that dimension.

Here the answer is no.  At least in the case of 4 qubits, there is
another type of state that also achieves the maximum nonproduct
stabilizer dimension.
The 4-qubit GHZ state has a 3-dimensional stabilizer, but so, for example,
does the 4-qubit state
\[
\ket{\psi} = \ket{0011} +   \ket{0101}
        - 2 \ket{0110} - 2 \ket{1001}
        +   \ket{1010} +   \ket{1100} ,
\]
whose stabilizer subalgebra consists of a single, 4-qubit $su(2)$ block.

Another example of a 4-qubit state with maximum nonproduct stabilizer
dimension, and a single 4-qubit $su(2)$ block is the state
\begin{align*}
\ket{M_4} &= \frac{1}{\sqrt{6}} [
  \ket{0011} + \ket{1100} + \omega (\ket{1010} + \ket{0101}) \\
  & \hspace{2cm}    + \omega^2 (\ket{1001} + \ket{0110}) ] ,
\end{align*}
with $\omega = \exp(2 \pi i/3)$.
This state was studied by A. Higuchi and others in
\cite{higuchi00} and \cite{brierley07}.
Among 4-qubit states, it has a local maximum
of average two-qubit bipartite entanglement, and has
the highest known value of this quantity.
We take this as another indication that enlarged stabilizers
indicate interesting states.

Note that dimension 4 is the only dimension in which a
state with an
$su(2)$ block can achieve the maximum nonproduct stabilizer dimension.
From the proof to Theorem \ref{maxdimnosingletnosingle},
we see that for nonproduct states with qubits outside $su(2)$ blocks,
the maximum stabilizer dimension is $n - p - 1$, so to achieve
the maximum nonproduct stabilizer dimension of $n-1$, these states
cannot have any $su(2)$ blocks.
For nonproduct states with all qubits in $su(2)$
blocks, Theorem \ref{thmnonsing} gives the maximum stabilizer dimension
to be $n - p$, which can only equal $n-1$ if $p=1$.  So, to achieve
maximum nonproduct stabilizer dimension with all qubits in $su(2)$ blocks,
there must be exactly one $su(2)$ block.  But one $su(2)$ block
for $n=6$ or higher gives a stabilizer dimension of 3 rather than $n-1$.

In light of these observations we make the following conjecture.

\begin{conjecture}
For $n = 3$ and $n \geq 5$, the generalized $n$-qubit GHZ state
and its LU-equivalents
are the only nonproduct states that achieve the maximum nonproduct
stabilizer dimension.
\end{conjecture}

\section{Acknowledgments}

The authors thank the National Science Foundation
for their support of this work through NSF Award No. PHY-0555506.



\end{document}